\documentclass[11pt,a4paper]{article}

\usepackage{footnote}

\usepackage[T1]{fontenc}
\usepackage[bitstream-charter]{mathdesign}

\usepackage[latin1]{inputenc}
\usepackage{amsmath}
\usepackage{cite}

\usepackage{xcolor}
\definecolor{bl}{rgb}{0.0,0.2,0.6}

\usepackage{graphicx}
\usepackage{sectsty}
\usepackage[compact]{titlesec}
\allsectionsfont{\color{bl}\scshape\selectfont}

\makeatletter
\def\printtitle{%
    {\color{bl} \centering \huge \sc \textbf{\@title}\par}}
\makeatother

\title{\\ \large \vspace*{-10pt} Exact Solutions of the Morse-like Potential, Step-Up and Step-Down Operators via Laplace Transform
Approach\vspace*{10pt}}

\makeatletter
\def\printauthor{%
    {\centering \small \@author}}
\makeatother

\author{%
    Altuð Arda \\
    arda@hacettepe.edu.tr \\
    Ramazan Sever \\
    sever@etu.edu.tr \\
    \vspace{20pt}
    }

\usepackage{fancyhdr}
\usepackage{lastpage}
\usepackage[runin]{abstract}
\abslabeldelim{\quad}
\catcode`ð=\active
 \defð{\u{g}}
 \catcode`Ð=\active
 \defÐ{\u{G}}
 \catcode`Ý=\active
 \defÝ{\. I}
 \catcode`ö=\active
 \defö{\"{o}}
 \catcode`Ö=\active
 \defÖ{\"O}
 \catcode`ü=\active
 \defü{\"{u}}
 \catcode`Ü=\active
 \defÜ{\"{U}}
 \catcode`Þ=\active
 \defÞ{\c{S}}
 \catcode`þ=\active
 \defþ{\c{s}}
 \catcode`ý=\active
 \defý{{\i}}
 \catcode`ç=\active
 \defç{\d{c}}
 \catcode`Ç=\active
 \defÇ{\d{C}}

\makesavenoteenv{tabular}
\begin{document}

\printtitle

\printauthor

\begin{abstract}
We intend to realize the step-up and step-down operators of the
potential $V(x)=V_{1}e^{2\beta x}+V_{2}e^{\beta x}$. It is found
that these operators satisfy the commutation relations for the
SU(2) group. We find the eigenfunctions and the eigenvalues of the
potential by using the Laplace transform approach to study the Lie
algebra satisfied the ladder operators of the potential under
consideration. Our results are similar to the ones obtained for
the Morse potential ($\beta \rightarrow -\beta$).
\end{abstract}

\section{Introduction}
Searching the exact solutions of the non-relativistic and
relativistic wave equations, \textit{i.e.}, Schrödinger equation
(SE), Klein-Gordon equation and Dirac equation, has became an
important part from the beginning of quantum mechanics \cite{sf}
and also in the view of the atomic and nuclear physics
\cite{lzy,csj,lj,ada1,ada2,si1,si2,vmv,oa}.

In this manner, the factorization method by which the creation and
annihilation operators of some potentials under consideration
could be obtained is a powerful tool to get the exact solutions of
some solvable potentials \cite{li,ybd}. This approach has been
received much attention in order to search the exact solutions in
the non-relativistic domain \cite{gh}. Bessis and co-workers have
extended the Schrödinger-Infeld-Hull factorization method
\cite{li,es} to the case, known as perturbed ladder operator
method, where the perturbed eigenvalues and eigenfunctions are
written in terms of the unperturbed physical system \cite{nb}.
Dong and co-workers have also used the factorization method with a
different point from the "old" one, namely, the ladder operators
under consideration can be constructed in terms of the physical
variable (\textit{i.e.}, without using an auxiliary nonphysical
variable) to obtain the dynamical group for different types of
potentials \cite{sd1,sd2,sd3,sd4,sd5,sd6,sd7,sd8}.

In this work, our aim is twofold. Firstly, we compute the
eigenvalues and the corresponding eigenfunctions of the
exponential-type potential, named as the Type-III potential
\cite{wjt}, by using the Laplace transform approach (LTA) which is
a economic method to obtain the exact solutions of the SE by
reducing it into a first-order differential equation
\cite{czd,gc1,gc2}. Secondly, we intend to search the raising and
lowering operators of this potential and give briefly the Lie
algebra of the commutators which falls into the SU(2) group. It is
seen that our results are similar to the ones obtained for the
Morse potential ($\beta \rightarrow -\beta$) \cite{sd2}.

\section{Energy Spectrum}

The one dimensional, time-independent Schrödinger equation is
written for a particle subjected to the potential under
consideration
\begin{eqnarray}
\frac{d^2\phi_{n}(x)}{dx^2}+\left\{ME_{n\ell}-MV_{1}e^{2\beta
x}-MV_{2}e^{\beta x}\right\}\phi_{n}(x)=0\,,
\end{eqnarray}
where $M=2m/\hbar^2$, $m$ is the mass and $E$ is the energy of the
particle.

Changing the variable to $z=e^{\beta x}$ in Eq. (1) gives the
following equation defined in an interval $z\in [0,\infty]$
\begin{eqnarray}
\frac{d^2\phi_{n}(z)}{dz^2}+\frac{1}{z}\frac{d\phi_{n}(z)}{dz}+\left\{-a^2_{1}-\frac{a^2_{2}}{z}-\frac{\varepsilon^2}{z^2}\right\}\phi_{n}(z)=0\,,
\end{eqnarray}
where
\begin{eqnarray}
a^2_{1}=\frac{MV_{1}}{\beta^2}\,\,;\,\,a^2_{2}=\frac{MV_{2}}{\beta^2}\,\,;\,\,-\varepsilon^2=\frac{ME_{n\ell}}{\beta^2}\,.
\end{eqnarray}

In order to get an equation having a suitable form for applying
the Laplace transform approach, we define a wave function
$\phi_{n}(z)=z^{\kappa}\varphi_{n}(z)$ which gives
\begin{eqnarray}z\frac{d^2\varphi_{n}(z)}{dz^2}-(2\varepsilon+1)\frac{d\varphi_{n}(z)}{dz}+\left\{-a^2_{2}-a^2_{1}z\right\}\varphi_{n}(z)=0\,,
\end{eqnarray}
where we set $\kappa=-\varepsilon$ to obtain a finite wave
function when $z \rightarrow \infty$.

By using the Laplace transform defined as \cite{mrs}
\begin{eqnarray}
\mathcal{L}\left\{\varphi(z)\right\}=f(t)=\int_{0}^{\infty}dze^{-ty}\varphi(z)\,,
\end{eqnarray}
Eq. (4) reads
\begin{eqnarray}
\left(t^2-a^2_{1}\right)\frac{df(t)}{dt}+\left[\left(\varepsilon+1\right)t
+a^2_{2}\right]f(t)=0\,,
\end{eqnarray}
which is a first-order ordinary differential equation and its
solution is simply given
\begin{eqnarray}
f(t)\sim\left(t+a_{1}\right)^{-\left(2\varepsilon+1\right)}
\left(\frac{t-a_{1}}{t+a_{1}}\right)^{-\frac{a^2_{2}}{2a_{1}}-\frac{2\varepsilon+1}{2}}\,,
\end{eqnarray}
In order to obtain a single-valued wave functions, it should be
\begin{eqnarray}
-\frac{a^2_{2}}{2a_{1}}-\frac{2\varepsilon+1}{2}=n\,\,\,(n=0, 1,
2, \ldots)
\end{eqnarray}
Using this condition and expanding Eq. (7) into series, we obtain
\begin{eqnarray}
f(t)\sim\sum_{k=0}^{n}\frac{(-1)^{k}n!}{(n-k)!k!}\,(2a_{1})^{k}(t+a_{1})^{-(2\varepsilon+1)-k}\,,
\end{eqnarray}
To obtain the solution of Eq. (4) we use the inverse Laplace
transformation \cite{mrs} and get
\begin{eqnarray}
\varphi_{n}(z)\sim
z^{2\varepsilon}e^{-a_{1}z}\sum_{k=0}^{n}\frac{(-1)^{k}n!}{(n-k)!k!}\,\frac{\Gamma(2\varepsilon+1)}{\Gamma(2\varepsilon+1+k)}\,(2a_{1}z)^{k}\,,
\end{eqnarray}
which gives
\begin{eqnarray}
\phi_{n}(z)=N_{n}z^{\varepsilon}e^{-a_{1}z}\sum_{k=0}^{n}\frac{(-1)^{k}n!}{(n-k)!k!}\,\frac{\Gamma(2\varepsilon+1)}{\Gamma(2\varepsilon+1+k)}\,(2a_{1}z)^{k}\,.
\end{eqnarray}
By using the following definition of the hypergeometric functions
\cite{ma}
\begin{eqnarray}
_{1}F_{1}(-n,\sigma,\xi)=\sum_{m=0}^{n}\frac{(-1)^{m}n!}{(n-m)!m!}\frac{\Gamma(\sigma)}{\Gamma(\sigma+m)}\xi^{m}\,,
\end{eqnarray}
and writing the hypergeometric functions in terms of Laguerre
polynomials as
$L_{n}^{p}(\xi)=\frac{\Gamma(n+p+1)}{n!\Gamma(p+1)}\,_{1}F_{1}(-n,p+1,\xi)$
\cite{ma}, we obtain the eigenfunctions as
\begin{eqnarray}
\phi_{n}(z)=N_{n}z^{\varepsilon}e^{-a_{1}z}\,\frac{n!\Gamma(2\varepsilon+1)}{\Gamma(n+p+1)}\,L_{n}^{2\varepsilon}(2a_{1}z)\,.
\end{eqnarray}
Using the normalization condition given as
$\int_{-\infty}^{\infty}\left|\phi(x)\right|^2dx=1$ the normalized
eigenfunctions are written
\begin{eqnarray}
\phi_{n}(z)= const.\,\sqrt{\frac{(2\varepsilon)^2
n!}{\Gamma(n+2\varepsilon+1)}\,}\,z^{\varepsilon}e^{-a_{1}z}\,L_{n}^{2\varepsilon}(2a_{1}z)\,,
\end{eqnarray}
where used [29]
\begin{eqnarray}
\int_{0}^{\infty}t^{\alpha-1}e^{-\delta t}L_{m}^{\lambda}(\delta
t)L_{n}^{\beta}(\delta
t)&=&\frac{\delta^{-\alpha}\Gamma(\alpha)\Gamma(n-\alpha+\beta+1)\Gamma(m+\lambda+1)}{m!n!\Gamma(1-\alpha+\beta)\Gamma(1+\lambda)}\nonumber\\
&\times&\,_{3}F_{2}(-m,\alpha,\alpha-\beta;-n+\alpha-\beta,\lambda+1;1)\,.
\end{eqnarray}
It is worth to say that '$const.$' in Eq. (14) includes some
factors related with hypergeometric function $\left(\,_{3}F_{2}(a,
b, c;r, s;1) \right)^{-1/2}$ and the parameter $a_{1}$ coming from
Eq. (15).

The requirement given in Eq. (8) and using the parameters in Eq.
(3), we find the energy eigenvalues of the exponential-type
potential as
\begin{eqnarray}
E_{n\ell}=-\frac{\beta^2}{4M}\left(2n+1+\frac{M}{\beta}\,\frac{V_{2}}{\sqrt{V_{1}\,}}\right)^2\,.
\end{eqnarray}
which gives the energy spectra of the Morse potential for $\beta
\rightarrow -\beta$. We give numerical energy eigenvalues of the
potential obtained from Eq. (16) for two diatomic molecules,
namely, for $H_2$ and $LiH$ molecule. For completeness, we also
summarize the eigenvalues for the Morse potential ($E_{M}$) by
setting $\beta \rightarrow -\beta$ in the same equation. The
parameter values of the molecules used here are as follows:
$D=4.7446$ eV, $r_{0}=0.7416 \AA$, $m=0.50391$ amu, $\alpha=\beta
r_{0}=1.440558$,
$E_{0}=\hbar^2/(mr^2_{0})=1.508343932\times10^{-2}$ eV for $H_2$
and $D=2.515287$ eV, $r_{0}=1.5956 \AA$, $m=0.8801221$ amu,
$\alpha=1.7998368$, $E_{0}=1.865528199\times10^{-3}$ eV for $LiH$
molecule \cite{in}. It is seen that being the potential parameter
$\beta$ positive causes to decrease the energy eigenvalues while
they increase in the case of the Morse potential.
\section{Step-up and Step-down Operators}
In this section, we tend to create briefly the ladder operators of
the potential satisfying the following eigenvalue equation
\begin{eqnarray}
\hat{L}_{\pm}\phi_{n}(z)=\ell_{\pm}\phi_{n}(z)\,,
\end{eqnarray}
where $\hat{L}_{+}$ is the step-up operator with eigenvalue
$\ell_{+}$ and $\hat{L}_{-}$ is the step-up operator with
eigenvalue $\ell_{-}$ and having the form
\cite{sd1,sd2,sd3,sd4,sd5,sd6,sd7,sd8}
\begin{eqnarray}
\hat{L}_{\pm}=f_{\pm}(z)\frac{d}{dz}+g_{\pm}(z)\,.
\end{eqnarray}

In order to get the step-down operator, we look for the acting of
the differential operator $d/dz$ on the eigenfunctions
\begin{eqnarray}
\frac{d}{dz}\,\phi_{n}(z)=\,\frac{\varepsilon}{z}\,\phi_{n}(z)-a_{1}\phi_{n}(z)+N_{n}z^{\varepsilon}e^{-a_{1}z}\frac{d}{dz}L_{n}^{2\varepsilon}(2a_{1}z)\,,
\end{eqnarray}
where if we take into account the constrain $2\varepsilon=-2n-1+A$
and use the property of the Laguerre polynomials \cite{isg}
\begin{eqnarray}
x\frac{d}{dx}L_{n}^{\alpha}(x)=nL_{n}^{\alpha}(x)-(n+\alpha)L_{n-1}^{\alpha}(x)\,,
\end{eqnarray}
we obtain
\begin{eqnarray}
\left(-z\frac{d}{dz}-a_{1}z+n+\varepsilon\right)\phi_{n}(z)=(n+2\varepsilon)\frac{N_{n}}{N_{n-1}}\,\phi_{n-1}(z)\,,
\end{eqnarray}
which gives the step-down operator
\begin{eqnarray}
\hat{L}_{-}=\sqrt{\frac{\varepsilon+1}{\varepsilon}\,}\left(-z\frac{d}{dz}-a_{1}z+n+\varepsilon\right)\,,
\end{eqnarray}
with
\begin{eqnarray}
\ell_{-}=(-n+A-1)\sqrt{n(n+2\varepsilon+1)\,}\,.
\end{eqnarray}
where $A=-a^2_{2}/a_{1}$. The last equation shows that the
step-down operator destroys the ground state. Using the following
recursion relation of the Laguerre polynomials \cite{isg}
\begin{eqnarray}
x\frac{d}{dx}L_{n}^{\alpha}(x)=(n+1)L_{n+1}^{\alpha}(x)-(n+\alpha+1-x)L_{n}^{\alpha}(x)\,,
\end{eqnarray}
and inserting into Eq. (19) gives
\begin{eqnarray}
\left(z\frac{d}{dz}-z+a_{1}z+n+\varepsilon+1\right)\phi_{n}(z)=(n+1)\frac{N_{n}}{N_{n+1}}\,\phi_{n+1}(z)\,,
\end{eqnarray}
From the last equation we obtain the step-up operator as
\begin{eqnarray}
\hat{L}_{+}=\sqrt{\frac{\varepsilon-1}{\varepsilon}\,}\left(z\frac{d}{dz}-z+a_{1}z+n+\varepsilon+1\right)\,,
\end{eqnarray}
with
\begin{eqnarray}
\ell_{+}=\sqrt{\frac{n+1}{-n+A+1}\,}\,.
\end{eqnarray}
The step-up operator in Eq. (26) annihilates the last bounded
state since for a such state is $\varepsilon=1$.

Finally we study the Lie algebra associated to the operators
$\hat{L}_{\pm}$ to construct the commutator of them with the help
of Eqs. (22) and (26):
\begin{eqnarray}
[\hat{L}_{+},\hat{L}_{-}]\phi_{n}(z)=\ell_{0}\phi(z)\,
\end{eqnarray}
where the eigenvalue
\begin{eqnarray}
\ell_{0}=2n+2-A\,,
\end{eqnarray}
which makes it possible to construct the operator
\begin{eqnarray}
\hat{L}_{0}=2\hat{n}+2-A\,.
\end{eqnarray}
These three operators satisfy the following Lie algebra
\begin{eqnarray}
[\hat{L}_{+},\hat{L}_{-}]=\hat{L}_{0}\,\,;[\hat{L}_{-},\hat{L}_{0}]=\hat{L}_{-}\,\,;[\hat{L}_{0},\hat{L}_{+}]=\hat{L}_{+}\,.
\end{eqnarray}
which correspond to the SU(2) group of the potential that means
the potential under consideration has the same group of the Morse
potential \cite{sd2}.

\section{Conclusion}
We have obtained the ladder operators of the Type-III potential
and commutation relations which correspond to the SU(2) group
which is also correspond to the ones of the Morse potential. To
achieve this aim, we have solved the Schrödinger equation for the
potential under consideration by using the Laplace transform
approach to find the eigenfunctions and eigenvalues. We have also
obtained the energy values of the Morse potential by setting
$\beta \rightarrow -\beta$ and summarized our numerical results
obtained for two diatomic molecules in Table 1.

\section{Acknowledgments}
This research was partially supported by the Scientific and
Technical Research Council of Turkey.


\newpage

\begin{table}[htp]
\caption{Energy eigenvalues of the exponential-type and Morse
potentials for different values of $n$ in $eV$ ($V_{1}=D,
V_{2}=2D$).}
\begin{tabular}{@{}ccccccc@{}}
&\multicolumn{3}{c}{$H_{2}$} &\multicolumn{3}{c}{$LiH$}
\\ \cline{2-4} \cline{5-7}
$n$ & $E_{n\ell}<0$  & $E_{M}<0$ & Ref. [30] & $E_{n\ell}<0$ & $E_{M}<0$ & Ref. [30]  \\
0 & 5.02101 & 4.47601 & 4.47601  & 2.60322 & 2.42886 & 2.42886 \\
2 & 6.20491 & 3.47992 & 3.47991  & 2.97007 & 2.09828 & 2.09827 \\
4 & 7.51402 & 2.60903 & 2.60902  & 3.36109 & 1.79186 & 1.79186 \\
10 & 12.1926 & 0.74759 &  0.74759 & 4.67918 & 1.01766 & 1.01765 \\
\end{tabular}
\end{table}

\end{document}